\documentclass[a4paper,11pt]{article}
\pdfoutput=1 % if your are submitting a pdflatex (i.e. if you have
\usepackage{jcappub} % for details on the use of the package, please
\usepackage[T1]{fontenc} % if needed
\usepackage{amsmath}
\usepackage{amssymb}
\usepackage{siunitx}
\usepackage[section]{placeins}
\usepackage{float}
\usepackage{comment}
\usepackage{slashed}
\usepackage[normalem]{ulem}
\usepackage{braket}
\usepackage{comment}

\title{\boldmath Dark Sector extensions of the Littlest Seesaw in the presence of Primordial Black Holes}

\author[a]{Baradhwaj Coleppa}
\author[a]{Kousik Loho}
\author[a,b]{Sujay Shil}

\affiliation[a]{Indian Institute of Technology Gandhinagar, Gujarat 382055, India}
\affiliation[b]{Instituto de F\'isica, Universidade de S\~ao Paulo, R. do Mat\~ao 1371, 05508-090 S\~ao Paulo, Brazil}
\emailAdd{baradhwaj@iitgn.ac.in}
\emailAdd{kousik.loho@iitgn.ac.in}
\emailAdd{sujayshil1@gmail.com}

\abstract{The Littlest Seesaw model is a  very well motivated dark matter model. Here we consider an extension of that model with an additional scalar and an additional fermionic particle under the freeze-in scenario. Formation of black hole of a certain mass range at primordial times can act as an alternate production mechanism for the dark matter particles as it evaporates via Hawking radiation. Furthermore, the presence of primordial black holes with substantial energy density gives rise to non-standard cosmology which also modifies the freeze-in production. In this paper, we have investigated the extended Littlest Seesaw model under the freeze-in scenario in the presence of a primordial black hole for various interesting cases and constrained the parameter space accordingly. If the universe is primordial black hole dominated at any point in the evolution of the universe, we find that the final relic in that case is dominated mostly by the evaporation component for a high dark matter mass and by the freeze-in component for a  low dark matter mass.}

\begin{document}
\maketitle
\flushbottom
\section{Introduction}

Over the last few decades, various experiments have established the Standard Model (SM) of particle physics as the most successful theory of elementary particles. However, some observations can not be explained within the purview of the SM and thus inspire us to search for Beyond Standard Model (BSM) physics. Non-zero neutrino masses and observation of neutrino oscillations \cite{Maki:1962mu,Bilenky:1978nj,Bilenky:2016pep,SNO:2001kpb} is one of the crucial indications of the existence of BSM physics. Over the years, several attempts have been made to account for the nonzero neutrino mass and oscillations, seesaw models being the most popular among them. The simple seesaw models contain three right-handed neutrinos in addition to the SM fields, a simpler version of which - dubbed the Littlest Seesaw (LS) model - has fewer parameters and two generations of Right-Handed Neutrinos (RHN) \cite{King:1999mb,King:2002nf}. In the LS model, a particular choice of the Yukawa matrix can explain the present value of the neutrino mixing angles and mass square differences. %In addition to the non-zero neutrino masses, there are a few other very crucial observations that motivate BSM searches.

Another significant evidence for BSM physics is the existence of Dark Matter (DM) \cite{Clowe:2006eq,Sofue:2000jx,Metcalf:2003sz,Bartelmann:2010fz}, which accounts for almost one fourth of the universe's energy budget \cite{Planck:2018vyg,WMAP:2012nax}. Many models have been developed in the past to explain the existence of dark matter. Among these,  one appealing class of models that explains both the dark matter and neutrino oscillations is called the Neutrino portal dark matter model \cite{Falkowski:2009yz,GonzalezMacias:2015rxl,Blennow:2019fhy,Bandyopadhyay:2020qpn,Bian:2018mkl,Coy:2021sse,Berlin:2018ztp,Borah:2021pet,Becker:2018rve}. These models have been extensively studied in the last few years. Typically in these models, there exist heavy RHNs $N_{R}$, which interact with the SM sector through the usual Yukawa type interactions $y_{N}\overline{L}\tilde{H}N_{R}$. This Yukawa interaction, along with the Majorana mass term of the heavy RHN, generates a tiny active neutrino mass by the so-called Type-I seesaw mechanism \cite{Minkowski:1977sc,Yanagida:1979as,Gell-Mann:1979vob,Mohapatra:1980yp}. The Dark Sector (DS) contains a singlet scalar field $\phi$ and a Dirac fermion $\chi$, which are connected to the SM through an interaction with the RHNs via the term $y_{DS}N_{R}\chi\phi$. For Yukawa couplings of strength comparable to that of the weak interaction, these particles of appropriate mass can act as the Weakly Interacting Massive Particle (WIMP) dark matter \cite{Bertone:2004pz,Bergstrom:2009ib,Arcadi:2017kky} as suggested by the so-called WIMP miracle \cite{Bauer:2017qwy}. However, those models are strongly constrained by various direct detection experiments \cite{XENON:2018voc,LUX:2016ggv,PandaX-II:2017hlx} which has motivated physicists to look for scenarios with smaller couplings that can survive these bounds. In such scenarios, the DM never thermalises with the cosmic soup and is produced in the early universe via out of equilibrium scattering and decay processes of the particles that are in equilibrium with the cosmic soup. This is the framework of the Feebly Interacting Massive Particle (FIMP) dark matter scenario \cite{Hall:2009bx,Bernal:2017kxu}. A few recent studies \cite{Chianese:2018dsz,Chianese:2019epo,Chianese:2020khl} have considered a minimal version of Type-I seesaw (i.e. the LS model) in the neutrino portal dark matter model by considering only two generations of RHNs where the possibility of the FIMP scenario has been explored.

 In most of the studies of neutrino portal dark matter the standard cosmological scenario is considered where the post inflation universe is dominated by radiation until matter took over. However, it is also interesting to study various non-standard cosmological scenarios in the context of the DM problem as is evident in the literature \cite{Cosme:2020mck,Arias:2019uol,Barman:2021ifu,Bernal:2018kcw,Bernal:2019mhf,DEramo:2017ecx,Arcadi:2020aot,Baldes:2020nuv,Bernal:2020ili,Dai:2009hx,Fujita:2014hha,Khlopov:2004tn,Chaudhuri:2020wjo,Morrison:2018xla,Hooper:2019gtx,Bernal:2020kse,JyotiDas:2021shi,Gondolo:2020uqv,Masina:2020xhk,Borah:2022vsu,Sandick:2021gew}. Among many such scenarios much interest has been shown in particular to the Primordial Black Holes (PBH) \cite{Baldes:2020nuv,Bernal:2020ili,Dai:2009hx,Fujita:2014hha,Khlopov:2004tn,Chaudhuri:2020wjo,Morrison:2018xla,Hooper:2019gtx,Bernal:2020kse,JyotiDas:2021shi,Gondolo:2020uqv,Masina:2020xhk,Borah:2022vsu,Sandick:2021gew} especially since the recent detection of Gravitational Waves (GW). Investigating the effects of an early matter dominated era on the DM phenomenology as opposed to the standard cosmological case has risen as an interesting prospect. PBH can form in the early Universe from density fluctuations but should evaporate away before Big-Bang Nucleosynthesis (BBN). One of the important aspects of studying PBH in the early Universe is that the mass of the PBH is small enough to evaporate into all the particles by Hawking radiation. So, Hawking radiation from PBH can be a source of dark matter origin in the early Universe. Another important aspect of the presence of PBH in the early Universe is that it can affect the Hubble expansion rate and change the standard cosmological scenario. The goal of the present work is to understand DM phenomenology with a PBH source in a model-specific context and also to understand how this scenario plays out in conjunction with other familiar avenues like the freeze-in mechanism. Specifically, we study a modified Type-I seesaw model and analyze the possibility of neutrino portal FIMP dark matter in  the presence of PBH. We explore the relevant parameter space to accommodate both the FIMP dark matter and the dark matter produced from PBH considering the present neutrino bounds.

The paper is organized as follows: in Sec.~\ref{sec:PBH}, the relevant background regarding PBH is briefly discussed and in Sec.~\ref{sec:model}, we describe our model which contains both the Littlest Seesaw and the dark sector (DS) components. We move on to the  discussion of the DM phenomenology in Sec.~\ref{sec:DMpheno} before concluding in Sec.~\ref{sec:conc}.

%%%%%%%%%%%%%%%%%%%%%%%%%%%%%%%%

\section{Primordial Black Hole Evaporation}
\label{sec:PBH}
In an example of a nonstandard scenario of cosmology, PBH can form due to the collapse of density fluctuations during an early radiation dominated era with initial temperature $T_{in}$. Just after formation, its initial mass ($M_{in}$) is given in terms of the energy density of radiation ($\rho_{R}$) by \cite{Carr:2009jm,Carr:2020gox,Bernal:2020bjf}
\begin{equation}
M_{in}=M(T_{in})=\frac{4\pi}{3}\gamma\frac{\rho_{R}(T_{in})}{H^{3}(T_{in})}; \ \ \ \ \rho_{R}(T)=\frac{\pi^{2}}{30}g_{*}(T)T^{4},
\end{equation}
where $\gamma=0.2$, $g_{*}(T)$ is the number of SM degrees of freedom as a function of temperature, and $H$ is the Hubble parameter. For a Schwarzschild black hole, the black hole temperature $T_{\rm BH}$ is related to its mass $M$ via \cite{Hawking:1974rv,Hawking:1975vcx}
\begin{align} \label{eq:TBH}
    T_{\rm BH}=\frac{M_p^2}{ M},
\end{align}
where $M_p$ is the reduced Planck mass. As the black hole evaporates, it produces every kind of particle through Hawking radiation. The rate of mass loss of the black hole is given by the expression \cite{MacGibbon:1990zk,MacGibbon:1991tj,Perez-Gonzalez:2020vnz}
\begin{align}\label{eq:MEq}
	\frac{dM}{dt}&= -\sum_i\varepsilon_i(M)\frac{M_p^4}{M^2},
\end{align}
where $\varepsilon_i$ depends on the particle species $i$ as well as the PBH mass and the $i$ is summed over to account for all species of particles. 

The presence of a black hole at primordial times modifies the cosmological history of the universe and thus has the potential to perturb  the evolution of the dark sector not only by injecting more dark matter particles into the spectrum via black hole evaporation but also through the modification of the Hubble parameter. The `non-standardness' of the cosmology is parameterised by defining a measure $\beta$, which is a fractional quantification of the initial black hole energy density:
\begin{equation}
\beta=\frac{\rho_{BH}^{in}}{\rho_{BH}^{in}+\rho_{R}^{in}}\sim\frac{\rho_{BH}^{in}}{\rho_{R}^{in}},
\end{equation}
and a very common rescaled version \cite{Cheek:2021cfe} is given by :
\begin{equation}
\label{eq:beta_p}
\beta^\prime=\sqrt{\gamma}\bigg(\frac{g_\star(T_{in})}{106.75}\bigg)^{-\frac{1}{4}}\beta\sim\sqrt{\gamma}\beta.
\end{equation}
%%%
The dominance of either component (black hole or radiation) is ascertained by comparing $\beta$ with a critical value $\beta_c$ which is given by \cite{Masina:2020xhk}
\begin{equation} 
\beta_c=\gamma^{-\frac{1}{2}}\bigg(\frac{\mathcal{G} g_{\star,H}(T_{BH})}{10640\pi}\bigg)^{\frac{1}{2}}\frac{M_{Pl}}{M},
\end{equation}
where $\mathcal{G}$ is the graybody factor and $M_{Pl}$ is the Planck mass. The BH evaporation can be traced by solving for the BH energy density ($\rho_{BH}$) from the equation \cite{Bernal:2020bjf}
\begin{equation}
\label{equ:bhevap}
\frac{d\rho_{BH}}{dt}+3H\rho_{BH}=\frac{\rho_{BH}}{M}\frac{dM}{dt},
\end{equation}
where $\frac{dM}{dt}$ is given in Eqn.~\ref{eq:MEq}.

Introduction of black hole(s) during primordial times is constrained by experimental observations as is to be expected. The excess radiation produced from PBH evaporation can modify successful Big Bang Nucleosynthesis (BBN) predictions unless it evaporates entirely before BBN (i.e. at temperatures higher than 4 MeV) \cite{Sarkar:1995dd,Kawasaki:2000en,Hannestad:2004px,deSalas:2016ztq,DEBERNARDIS2008192}. This can be translated into an upper bound on the initial mass of PBH : $M_{in}\lesssim 10^9$ g. Furthermore, there exists a lower bound on the initial mass from the upper bound on the inflationary scale: $M_{in}\geq0.1$ g \cite{Planck:2018jri}. In addition, GW can be produced from the density perturbations as a consequence of the presence of PBH. Hence, the parameter space has to be constrained accordingly for a monochromatic PBH mass distribution with the assumption that the energy of GW never overtakes the energy of the background universe: $\beta\leq 10^{-4}\big(\frac{10^9 \textrm{g}}{M_{in}}\big)^{\frac{1}{4}}$ \cite{Domenech:2020ssp,Papanikolaou:2020qtd,Bernal:2020bjf}.

The effect of PBH evaporation on the DM relic is typically described in terms of three very important parameters: the initial mass of the PBH ($M_{in}$), the mass of the DM candidate ($m_{DM}$), and $\beta$. In the PBH dominated region ($\beta>\beta_c$), the DM relic produced from PBH evaporation does not depend on $\beta$ but is a function of both $M_{in}$ and $m_{DM}$. In the radiation dominated regime ($\beta<\beta_c$) however, along with the $M_{in}$ and $m_{DM}$ dependence, the relic is also proportional to $\beta$. The value of $m_{DM}$ in comparison with the initial PBH temperature $T_{in}$ plays a crucial role in determining some of these dependences as described in \cite{Bernal:2020ili}. For a heavy DM candidate (i.e., $m_{DM}>T_{in}$), the PBH has to evaporate significantly via other low mass particles to raise its temperature (defined in Eqn.~\ref{eq:TBH}) to $m_{DM}$ before it can start emitting DM particles. A few of these features will be evident from our results in Sec.~\ref{sec:DMpheno}.

%%%%%%%%%%%%%%%%%%%%%%%%%%%%%%%%%%%%%%%%%%%%%%%%%%%%%%%%%%%%%%

\section{Model}
\label{sec:model}

In this work, we consider the simplest version of Type-I seesaw model \cite{Chianese:2018dsz} with two RHNs $N_{R_\beta}$, where $\beta=1,2$ is the generation index. In addition, there are two extra dark sector (DS) particles: a scalar $\phi$ and a fermion $\chi$.  The dark sector particles are singlets under the SM gauge group $SU(2)_{L}\times U(1)_{Y}$ and so are the RHNs. The model has an additional $Z_{2}$ symmetry under which the dark sector particles are odd and everything else including the SM particles is even - the transformation properties of the BSM particles under the symmetry groups of the model is shown in Table \ref{tab:sym}. 
\begin{table}[h!]
\centering
\begin{tabular}{|c|c|c|c|}
\hline
& $N_R$ & $\phi$ & $\chi$ \\ \hline \hline 
$SU(2)_L$ & {\bf 1} & {\bf 1} & {\bf 1} \\ \hline
$U(1)_Y$ & 0 & 0 & 0 \\ \hline  \hline
$Z_2$ & + & - & - \\ \hline
\end{tabular}
\caption{Transformation of the BSM particles under the different symmetries of the model.}
\label{tab:sym}
\end{table}

The Lagrangian of the model is written as
\begin{equation}
\mathcal{L} = \mathcal{L}_{\rm SM} + \mathcal{L}_{\rm seesaw} + \mathcal{L}_{\rm DS} + \mathcal{L}_{\rm portal}\ ,
\label{eq:lag}
\end{equation}
where
\begin{eqnarray}
\mathcal{L}_{\rm seesaw} & = & - Y_{\alpha\beta} \overline{L_L}_\alpha \tilde{H} N_{R\beta} - \frac12 M_{R}\overline{N^c_{R}} N_{R} +h.c.,\ \label{eq:lagNS} \\
\mathcal{L}_{\rm DS} & = & \overline{\chi}\left(i \slashed{\partial} - m_\chi \right)\chi +\frac{1}{2} \left|\partial_\mu \phi\right|^2 - m^2_\phi \left|\phi\right|^2 - V\left(\phi\right),\ \label{eq:lagDS} \\
\mathcal{L}_{\rm portal} & = & -y_{\rm DS} \phi \overline{\chi}N_{R} + h.c., \ \label{eq:lagPortal}
\end{eqnarray}
and $\tilde H=i\sigma_2H^\star$. The $\mathcal{L}_{\rm seesaw}$ dictates the Littlest Seesaw (LS) \cite{King:2015dvf} mechanism with two RHNs in this model. The $\mathcal{L}_{\rm DS}$ governs the interactions of the $\chi$ and $\phi$ particles with themselves. The $\mathcal{L}_{\rm portal} $, as can be seen from Eqn.~\ref{eq:lagPortal}, facilitates interaction of the DS particles with the heavy RHNs and thus provides a heavy neutrino portal for the DS particles to interact with the SM sector. However, in certain scenarios we will see that the RHNs themselves can also serve as DM candidates. In this work we would like to focus on this neutrino portal as the sole messenger between the SM and the DS sectors and hence we will not consider any possible interactions between the $\phi$ and the SM $H$. The explicit form of $V(\phi)$ need not be specified as long as it respects the $Z_2$ symmetry since we are not interested in the specific scalar sector interactions. Furthermore, we will work under the assumption that the $Z_2$ symmetry of the DS remains intact throughout the evolution of universe.

After Electroweak Symmetry Breaking (EWSB), the Higgs field develops a vacuum expectation value (vev) $v$ and consequently the Dirac mass term of the neutrino can be written as $m_D = \frac{v}{\sqrt 2} Y$. This minimal seesaw example with only two RHNs survives the constraints from physical neutrino observables only for a one texture-zero structure \cite{King:2013iva} case. The heavy neutrinos which are understood as the right handed atmospheric and solar components, couple to the flavour specific active neutrinos with a particular ratio \cite{King:2018fqh}. Hence, the Dirac mass matrix is written in the following fashion such that the structure is consistent with the reactor angle \cite{King:2015dvf} along with the other relevant parameters: 

\begin{equation}
m_D = \left(\begin{array}{lr} 0 &  b \, e^{i \frac{\eta}{2}} \\ a & 3b \, e^{i \frac{\eta}{2}} \\ a & b \, e^{i \frac{\eta}{2}} \end{array}\right). \,
\label{mD2}
\end{equation}
The Majorana mass matrix has been chosen to be diagonal:
\begin{equation}
M_R =  \left(\begin{array}{cc} M_{\rm R_1} &  0 \\ 0 & M_{\rm R_2} \end{array}\right).
\end{equation}
%%
%\Barath{For the future, pmatrix typsets better matrices than the array environment. No need to change anything here though.}
%
Using the Type-I seesaw formula 
\begin{equation}
m_{light}\sim m_DM_R^{-1}m_D^T, \qquad m_{heavy} \sim M_{R},
\end{equation}
 the light neutrino mass matrix can be written in this model as
\begin{equation}
m_\nu = m_a \left(\begin{array}{ccc} 0 & 0 & 0 \\ 0 & 1 & 1 \\ 0 & 1 & 1 \end{array}\right) + m_b \, e^{i \eta} \left(\begin{array}{ccc} 1 & 3 & 1 \\ 3 & 9 & 3 \\ 1 & 3 & 1 \end{array}\right),\,
\label{eq:numass}
\end{equation}
where $m_a$ and $m_b$ are given by
\begin{equation}
m_a = \frac{a^2}{M_{R_1}} \qquad {\rm and} \qquad m_b = \frac{b^2}{M_{R_2}}.
\label{mD3}
\end{equation}
The Yukawa matrix is accordingly 
\begin{equation}
Y = \sqrt{\frac{2 \,m_a\, M_{\rm R_1}}{v^2}} \left(\begin{array}{cc} 0 & 0 \\ 1 & 0 \\ 1 & 0\end{array}\right) + \sqrt{\frac{2 \,m_b\, M_{\rm R_2}}{v^2}} e^{i \frac{\eta}{2}} \left(\begin{array}{cc} 0 & 1 \\ 0 & 3  \\ 0 & 1\end{array}\right)\,.
\label{eq:yuk}
\end{equation}
The $m_{\nu}$ matrix has to be diagonalized further to get the mass eigenstates of the active neutrinos. After diagonalizing, let us denote the mass eigenvalues of the active neutrinos as $m_{1}$, $m_{2}$ and $m_{3}$. Considering $m_{1}=0$ with normal hierarchy, the atmospheric neutrino mass $m_{3}$ is dominantly controlled by $m_{a}$ and solar neutrino mass $m_{2}$ is dominantly controlled by $m_{b}$. 
To explain the neutrino masses and other oscillations data (the explicit equations that describe the relations can be found in Ref.~\cite{King:2016yvg}), the following benchmark values have been chosen:
\begin{equation}
m_a = 26.7~{\rm \,meV}\,, \qquad m_b = 2.7~{\rm \,meV}, \qquad {\rm and} \qquad \eta = \frac23\pi\ ,
\label{eq:nu-data}
\end{equation}
which are consistent with the experimental values of PMNS parameters as well as the light neutrino mass differences at $3\sigma$ range \cite{Esteban:2020cvm}. The  explicit values for our model are given in Table~\ref{tab:param}.

\begin{table}[h!]
\centering
\begin{tabular}{|c|c|c|c|c|c|c|}
\hline
$\theta_{13}$&$\theta_{12}$&$\theta_{23}$&$\delta_{CP}$&$m_1$&$m_2$&$m_3$ \\
 \hline 
\ang{8.67}&\ang{34.32}&\ang{45.78}&\ang{-86.63}&0 meV&8.64 meV&50.06 meV \\
 \hline

\end{tabular}
\caption{PMNS parameters and the active neutrino masses in the current model.}
\label{tab:param}
\end{table}

The relevant terms of the Lagrangian that take part in the freeze-in mechanism are outlined in the following equation in terms of the mass eigenstates:

\begin{equation}
\mathcal{L}\supset-(U_\nu^\dagger Y)_{ij} \nu_i\frac{h-iG^0}{\sqrt{2}}N_{R_j}+Y_{ij}l^+_i G^-N_{R_j} -y_{\rm DS} \phi \overline{\chi}N_{R} +h.c. ,
\end{equation}
where for the sake of simplicity we have assumed that both the heavy neutrinos couple to the DS with same strength (i.e., $y_{DS}$ is same for both the heavy neutrinos). The PMNS matrix $U_\nu$ has been applied accordingly to account for the flavour mixing in the active neutrino sector. The scalar field $h$ is identified with the SM Higgs and the Goldstone modes $G^0$ and $G^\pm$ are  eaten up by the weak gauge bosons via the Higgs mechanism.

%%%%%%%%%%%%%%%%%%%%%%%%%%%%%%%%%%%%%%%%%%%%%%%%%%%%%%%%%%%%%%

\section{Dark Matter Phenomenology}
\label{sec:DMpheno}

In this section, we will analyse the DM phenomenology in the presence of a monochromatic PBH distribution in two very distinct mass regions  for two interesting mass hierarchies in each case. Depending on the choice of the mass region and mass hierarchy, the two heavy neutrinos and/or the dark sector particles may serve as dark matter candidates. We choose to explore the freeze-in scenario in the context of a dark sector isolated from the thermal bath due to a very small coupling $y_{DS}$. The heavy neutrinos can work as DM portals or as the DM candidates themselves depending upon the scenario under consideration. The DM candidates in this case do not ever thermalise with the bath due to the inherently small couplings which is also responsible for the out of equilibrium production of the dark matter particles from the bath particles, and as the temperature of the bath cools down upto around the DM mass it ceases to produce DM anymore and freeze-in takes place. 

The evolution of the universe is modified in the presence of the PBH since it radiates all kinds of particles of the spectrum and in addition, the PBH energy density also has to be taken into account while computing the Hubble parameter. While this acts as an additional source of production for the DM particles, it also dilutes the relic via entropy injection. This effect can be drastically different in various regions of the parameter space. In what follows, we study two very different mass regions of the DM model and present our findings categorised into four example cases for ease of analysis. Further, for the sake of simplicity, we have chosen the two heavy neutrinos to be degenerate in mass ($m_{R_1}=m_{R_2}=m_N$)\footnote{Relaxing this assumption, would, of course, introduce additional decay channels which need to be taken into account for the DM phenomenology.} in all cases under consideration. In what follows, the choice of the masses of the DS particles along with the heavy neutrinos is primarily based on PBH evaporation. For a range of DM masses around $10-10^9$ GeV, PBH evaporation by itself can not produce the correct relic in the PBH dominated region due to BBN constraints as shown in \cite{Cheek:2021cfe}. Hence, we have chosen to explore masses much higher and much lower than this region.

\subsection{Case-I : High Mass Region with $\boldsymbol{m_\phi>m_\chi>m_N}$}
\label{subsec:case1}

In this scenario we choose the benchmark point $m_\phi= 10^{14}$ GeV, $m_\chi=10^{12}$ GeV, and $m_N=10^{10}$ GeV. To satisfy the sub-eV light neutrino masses, an $\mathcal{O}(10^{-3}-10^{-4})$ Yukawa coupling is required which results in thermalization of the heavy neutrinos. Furthermore, the $y_{DS}$ is chosen to be $\mathcal{O}(10^{-8})$ so that the dark sector is isolated and DM is produced via the freeze-in mechanism. Apart from PBH evaporation the production mechanism consists of annihilation of SM particles through $s$-channel heavy neutrino portals as well as the annihilation of the heavy neutrinos themselves via $t$ and $u$ channel diagrams as the dark sector particles take turns in mediating \cite{Chianese:2018dsz}. Additionally, the $\phi$ particles produced from the freeze-in and PBH evaporation decay into $\chi$ and $N$ as it is kinematically allowed because of the chosen hierarchy of masses. The heavy neutrinos also are washed out via SM decays, effectively leaving only the $\chi$ to form the DM relic.

The Boltzmann equations that dictate the evolution of the number densities of the DS particles by freeze-in mechanism are given as
\begin{equation}
\frac{dn_\chi^{FI}}{dt}+3Hn_\chi^{FI}=n_{\chi eq}^2\langle\sigma v\rangle_{\chi\chi\rightarrow NN}+n_{\chi eq}n_{\phi eq}\langle\sigma v\rangle_{\chi\phi\rightarrow SMSM}+2n_\phi^{FI}\langle\Gamma_{\phi\rightarrow\chi N} \rangle, 
\end{equation}
\begin{equation}
\frac{dn_\phi^{FI}}{dt}+3Hn_\phi^{FI}=n_{\phi eq}^2\langle\sigma v\rangle_{\phi\phi\rightarrow NN}+n_{\chi eq}n_{\phi eq}\langle\sigma v\rangle_{\chi\phi\rightarrow SMSM}-2n_\phi^{FI}\langle\Gamma_{\phi\rightarrow\chi N}\rangle.
\end{equation}
The SM state pairs mentioned in the above equation are $\nu h$, $\nu G^0$, and  $l^\pm G^\mp$ (where $l^\pm$ are the leptons and the $G$'s are the Goldstone modes of the scalar doublet $H$). 

A separate set of equations track the number densities of these particles produced from BH evaporation independently from their freeze-in counterparts:

\begin{equation}
\frac{dn_\chi^{BH}}{dt}+3Hn_\chi^{BH}=\frac{\rho_{BH}}{M}\frac{dN_{\chi}}{dt}+2n_\phi^{BH}\langle\Gamma_{\phi\rightarrow\chi N}\rangle, 
\label{equ:reg1_3}
\end{equation}

\begin{equation}
\frac{dn_\phi^{BH}}{dt}+3Hn_\phi^{BH}=\frac{\rho_{BH}}{M}\frac{dN_{\phi}}{dt}-2n_\phi^{BH}\langle\Gamma_{\phi\rightarrow\chi N}\rangle, 
\label{equ:reg1_4}
\end{equation}
where $\frac{dN_i}{dt}$ is the emission rate of species $i$ from the evaporating BH. The factor of 2 in front of the $\phi$ decay term on the right hand side of the above equations is because of two generations of heavy neutrinos. A few more relevant evolution equations are given in Appendix \ref{sec:equations} and also in Eqn. \ref{equ:bhevap}. It has to be noted that there are two temperatures involved which are the plasma temperature and the BH temperature. The thermal averaging has to be done accordingly depending upon the relevant production mechanism. In contrast, the thermal averaging for $\langle\sigma v\rangle$ is done always with respect to the plasma temperature. The definition of $\langle\sigma v\rangle$ employed in this paper follows that of Ref.~\cite{Chianese:2018dsz}.

\begin{figure}[h!]
\centering
\includegraphics[scale=0.55]{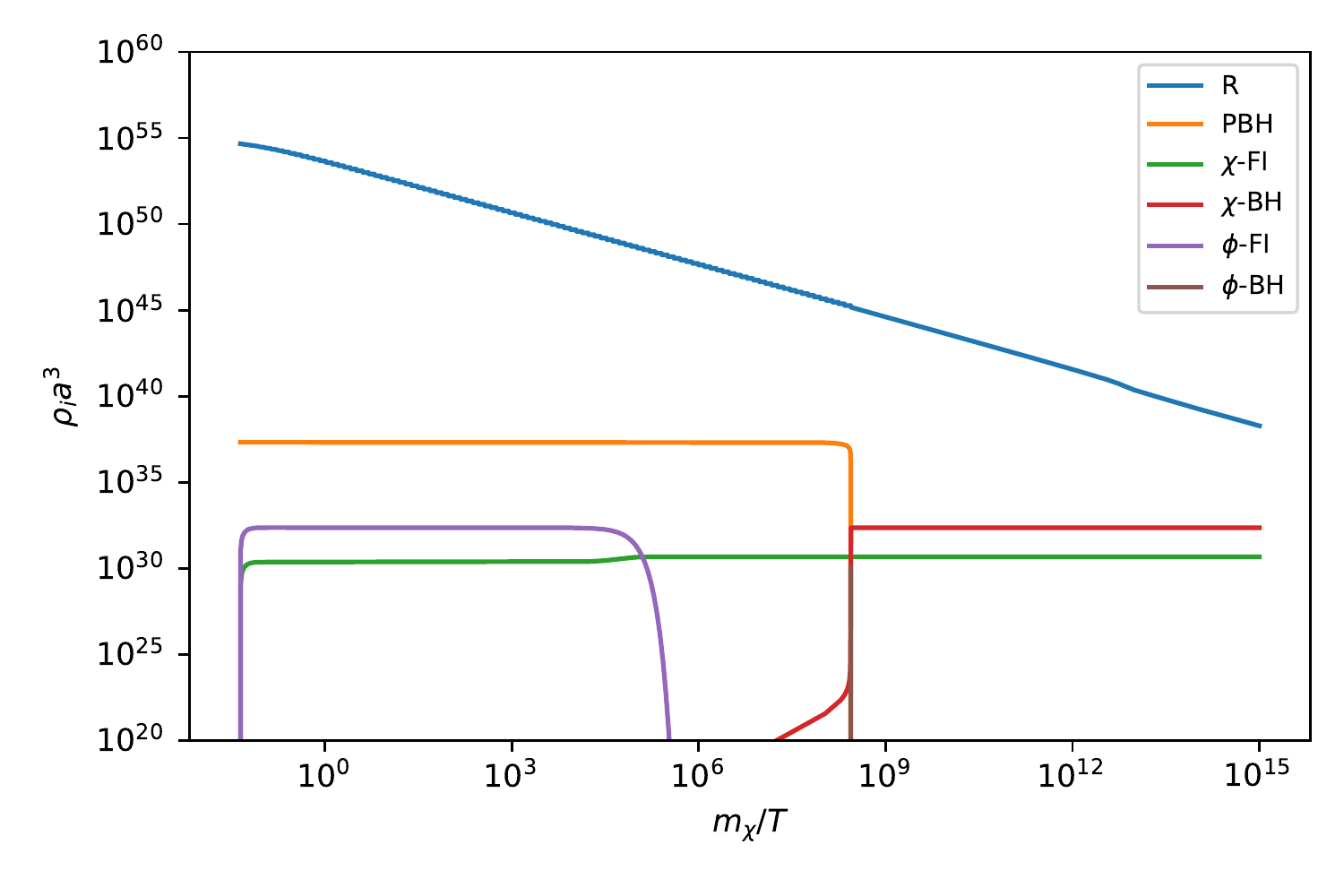}\includegraphics[scale=0.58]{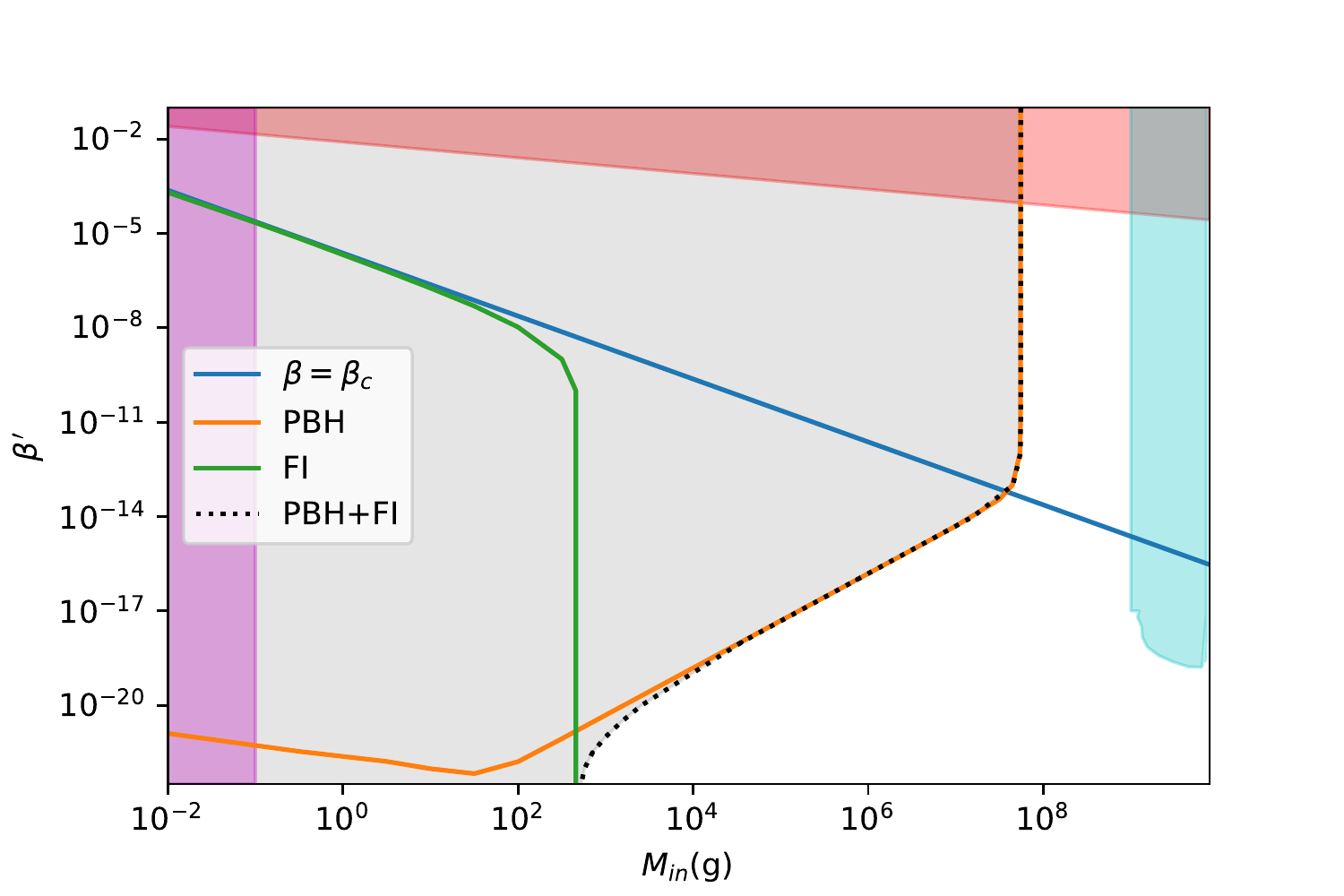}
\caption{ Energy density (times co-moving volume) evolution of various components of universe for benchmark points $M_{in}=10^{4.55}$ g and $\beta^\prime=10^{-18.0}$ where $\Omega h^2\sim0.12$ (left).  Relic contours ($\Omega h^2=0.120\pm0.001$) for Freeze-In (FI) and Primordial Black Hole (PBH) components individually as well as summed together along with various PBH constraints (right).}
\label{fig:relic1}
\end{figure}

We have solved the above mentioned Boltzmann equations numerically to trace the evolution of relevant individual components of the universe and determine  the final relic abundance\footnote{We have used the publicly available code ULYSSES \cite{Cheek:2021cfe,Cheek:2021odj} to model the PBH evaporation.}. Fig. \ref{fig:relic1} (left) shows the evolution of the co-moving energy densities (radiation is the exception since $\rho_{R}\sim a^{-4}$, where $a$ is the scale factor of the universe not to be confused with model parameter $a$ in Sec. \ref{sec:model}) of the relevant components of universe. In this particular example, one can clearly point out that the energy density of PBH never dominates over radiation at any point of its evolution. The radiation component shows a steady linear decline with the cooling of universe since $\rho_{R}a^3\sim a^{-1}\sim T$. Both $\phi$ and $\chi$ are produced via freeze-in at very high temperatures\footnote{In the high mass region particularly in this scattering dictated freeze-in scenario, freeze-in can predate the PBH formation which explains the immediate saturation of $\phi$-FI and $\chi$-FI in Fig. \ref{fig:relic1} (left) at high temperatures. In that case the existing number density of the freeze-in components at the time of PBH formation should be fed as initial conditions while solving the Boltzmann equations. However, that changes our final relic abundance by less than 1\% of the observed value and hence can be neglected.}. However, as $\phi$ decays into $\chi$ and a right handed neutrino, leaving no trace of frozen-in $\phi$ beyond $\frac{m_\chi}{T}=10^6$, the co-moving $\rho_{\chi-FI}$ is slightly raised in logarithmic scale by that production from $\phi$ decay. In addition to freeze-in, both $\phi$ and $\chi$ are also produced from BH evaporation. The sharp decay of PBH as displayed in Fig. \ref{fig:relic1} (left) escalates the particle production from BH evaporation. While the $\phi$ produced from BH evaporation decays immediately, $\chi$ produced from BH evaporation remains as a relic along with its freeze-in counterpart.

\FloatBarrier

The relic contours ($\Omega h^2=0.120\pm0.001$ as per Planck data \cite{Planck:2018vyg}) have been shown in Fig.~\ref{fig:relic1} (right) considering only the freeze-in, only the BH evaporation and also both of them put together. Furthermore, various constraints on PBH mentioned in Sec.~\ref{sec:PBH} are also superimposed on this plot: the two vertical shaded regions are excluded by BBN (light cyan) and inflation (light magenta) observations and also the one from GW (light red) data. The grey shaded region is excluded by overabundance. One important thing to observe is that in this scenario the freeze-in components do not dictate the relic in the PBH dominated region ($\beta>\beta_c$). Also, for very small values of $\beta$ in some parts of the radiation dominated region ($\beta<\beta_c$), the effect of PBH in cosmology, apart from particle production, becomes negligible which is reflected in the freeze-in relic as it becomes independent of $\beta$ in that region. However in this case the contour depends on the initial mass of the primordial black hole, i.e., the initial temperature of the early universe when the black hole forms and also the freeze-in process takes off. Hence, in the radiation dominated universe the frozen-in relic depends only on the initial temperature. Also, in the presence of a significant amount of early matter content, a fast expanding universe dilutes the relic and thus requires a high initial temperature (i.e., low PBH mass). In this case-I, the freeze-in production is driven by scattering processes which are typically less productive than the freeze-in production of case-II where it is driven by decay processes. Hence, typically a decay driven production (case-II) can overcome the entropy dilution more effectively as compared to a scattering driven production (case-I). This phenomena is reflected in the freeze-in contour as it can satisfy the relic only in the radiation dominated region in case-I (Fig \ref{fig:relic1} right). However, in case-II (Fig \ref{fig:relic2} right), we will see that the freeze-in relic can attain the observed value in the PBH dominated region as well. Dark matter production from PBH evaporation hinges on a very important criteria i.e. the BH temperature has to be greater than the mass of the dark matter particle in order to produce it. If the initial BH temperature is not as high as the mass of the DM particle, then it can only radiate other low mass particles. Radiating other particles however brings down the BH mass and thus raises the BH temperature (see Eqn.~\ref{eq:TBH}) allowing it to produce the massive DM particles. Depending upon whether or not the PBH can produce DM particles initially, the PBH relic in Fig.~\ref{fig:relic1} (right) has two slopes in the radiation dominated region. In the PBH dominated region however, the relic is of course independent of $\beta$ as expected \cite{Bernal:2020ili}. The total relic (i.e. FI+PBH) contour separates the under-abundant and over-abundant regions where a higher initial PBH mass typically leads to more entropy injection thus diluting the relic into under-abundance. Furthermore, we find the relevant velocity averaged cross sections to be too small for $\chi$ and $\phi$ to ever thermalise with the bath. Finally, we have checked to ascertain that such massive DM particles are indeed immune to hot dark matter constraints \cite{Baldes:2020nuv,Baur:2017stq}.

\subsection{Case-II : High Mass Region with $\boldsymbol{m_N>m_\phi=m_\chi}$}
\label{subsec:case2}

We now turn to the analysis of the benchmark point $m_\phi=m_\chi=10^{12}$ GeV. A heavy neutrino mass of $10^{14}$ GeV and corresponding $\mathcal{O}(10^{-1}-10^{-2})$ Yukawa couplings generate the sub-eV active neutrino masses. The out of equilibrium decay of thermalized heavy neutrinos into $\phi$ and $\chi$ with $y_{DS}=10^{-8}$ populates the dark sector in addition to the production from PBH evaporation. In order not to complicate the scenario further, a conscious choice has been made to keep the dark sector masses degenerate - if this condition is relaxed, one DS particle could decay into the other one and two SM states via an off-shell $N$. However, in our relatively simple scenario, both $\phi$ and $\chi$ serve as dark matter candidates.

The evolution of the DS particles is dominated by heavy neutrino decays as indicated in the following equations:
\begin{equation}
\frac{dn_\chi^{FI}}{dt}+3Hn_\chi^{FI}=2n_{Neq}\langle\Gamma_{N\rightarrow\phi\chi}\rangle, 
\end{equation}

\begin{equation}
\frac{dn_\phi^{FI}}{dt}+3Hn_\phi^{FI}=2n_{Neq}\langle\Gamma_{N\rightarrow\phi\chi}\rangle.
\end{equation}
The factor of 2 on the right hand side accounts for the two generations of heavy neutrinos.
The BH contribution also adds up in calculating the total relic - this can be determined by solving the following equations:

\begin{equation}
\frac{dn_\chi^{BH}}{dt}+3Hn_\chi^{BH}=\frac{\rho_{BH}}{M}\frac{dN_{\chi}}{dt},
\end{equation}

\begin{equation}
\frac{dn_\phi^{BH}}{dt}+3Hn_\phi^{BH}=\frac{\rho_{BH}}{M}\frac{dN_{\phi}}{dt}. 
\end{equation}

\begin{figure}[h!]
\centering
\includegraphics[scale=0.55]{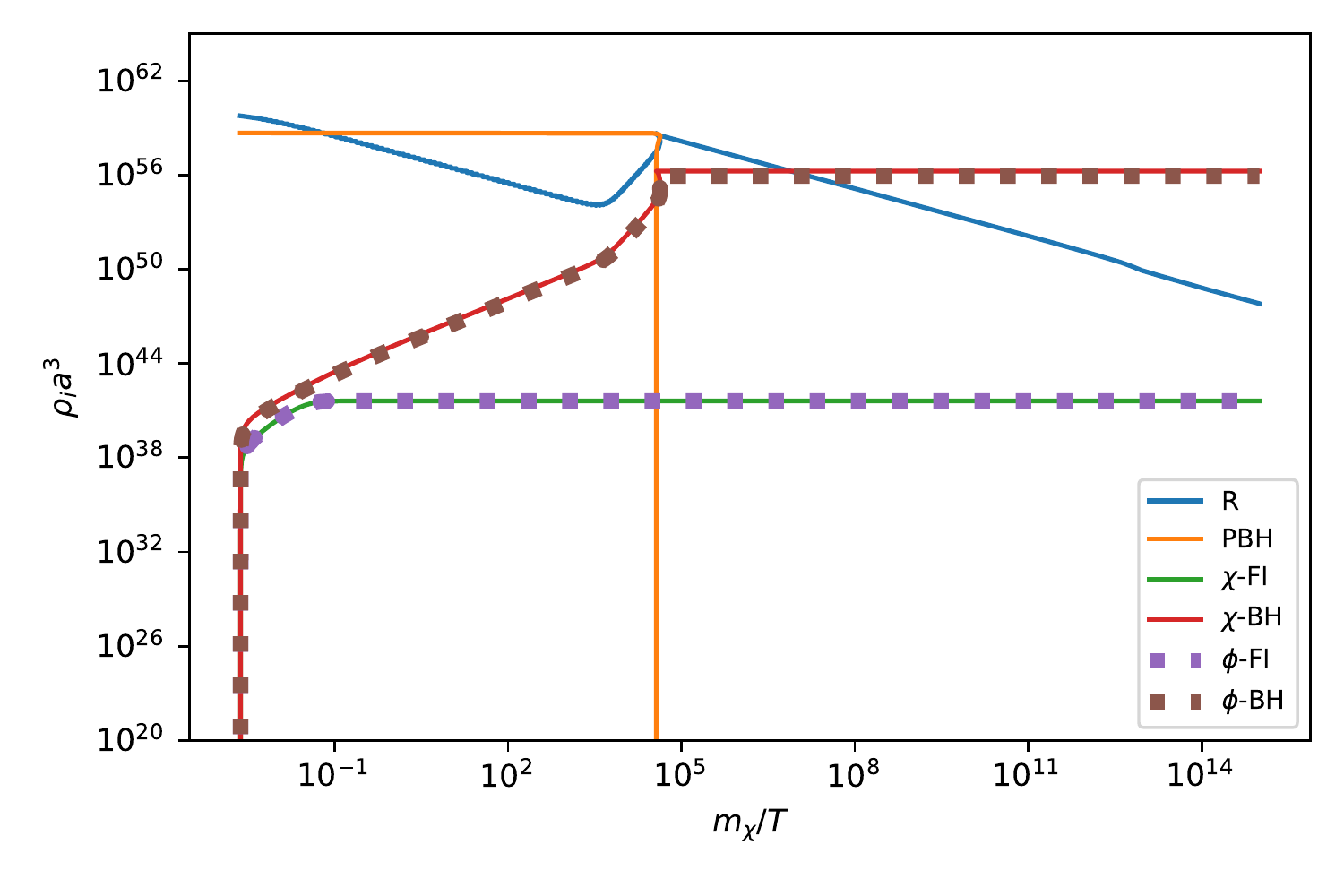}\includegraphics[scale=0.55]{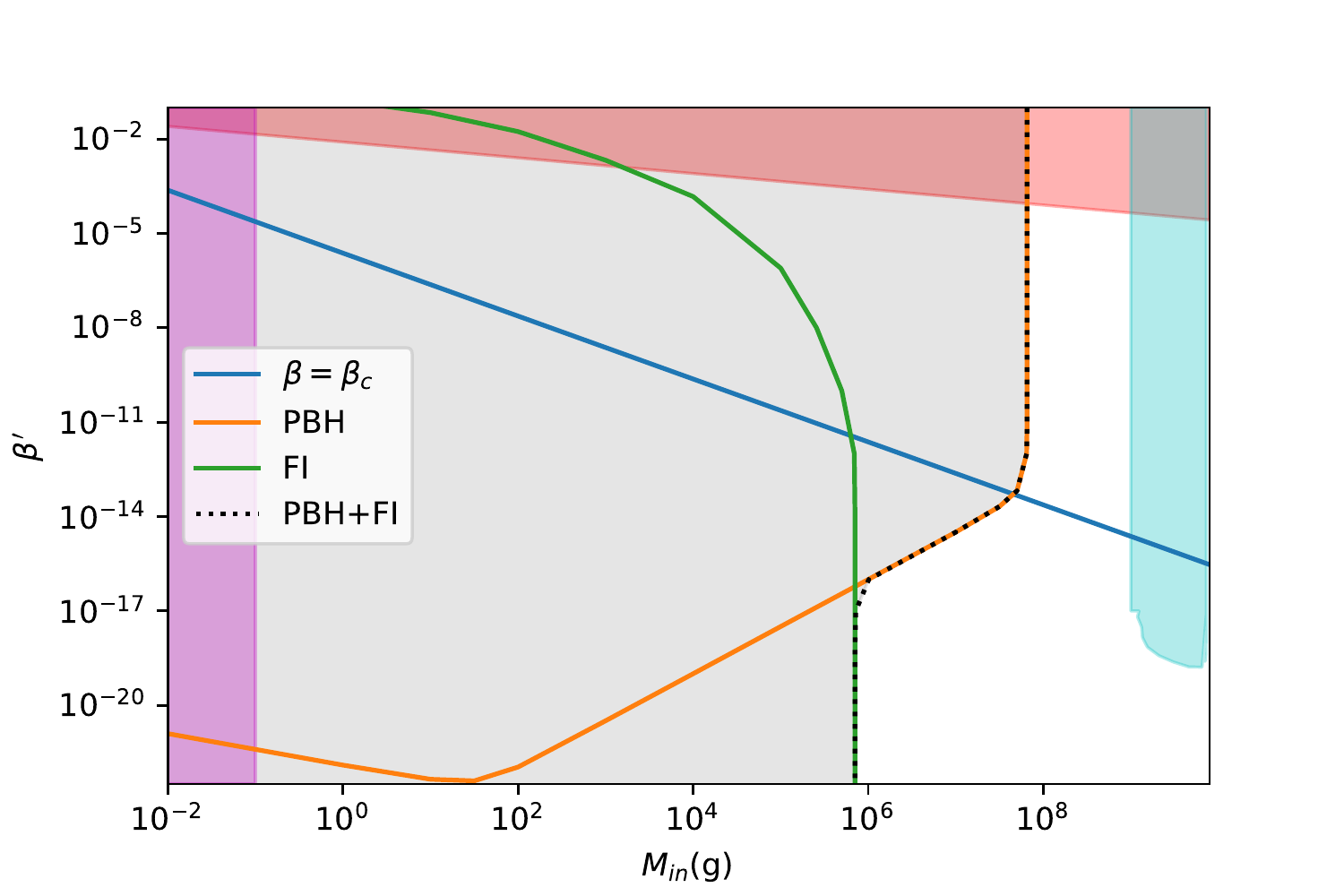}
\caption{ Evolution of various components of universe for benchmark points $M_{in}=10^{2.0}$ g and $\beta^\prime=10^{-1.765}$ where $\Omega_{FI} h^2\sim0.12$ (left).  Relic contours for Freeze-In (FI) and Primordial Black Hole (PBH) components individually as well as summed together along with various PBH constraints (right).}
\label{fig:relic2}
\end{figure}

The evolution of the co-moving energy densities of the degenerate dark sector particles in most parts mimic each other as has been shown in Fig.~\ref{fig:relic2} (left). The freeze-in takes place early enough through the heavy neutrino decay into $\chi$ and $\phi$. The component from PBH evaporation also plays its part in forming the relic. The volume multiplied energy densities of BH and radiation in this case are quite comparable and a region of PBH domination is evident in Fig.~\ref{fig:relic2} (left). One should also take note of the fact that entropy injection due to PBH evaporation results in the increase of radiation energy density before starting to decrease again once the PBH evaporation is complete.

\FloatBarrier

Even in this decay scenario with this chosen benchmark values the thermal contribution does not constitute a major portion of relic in the PBH dominated region as is evident in Fig.~\ref{fig:relic2} (right). The inference in this case is quite similar to that of case-I apart from a key difference in the freeze-in relic as mentioned in Sec.~\ref{subsec:case1}.

\subsection{Case-III : Low Mass Region with $\boldsymbol{m_\phi>m_\chi>m_N}$}
We now turn to analyzing the orthogonal parameter space wherein the masses are all small - in particular, we choose the following benchmark values: $m_\phi=0.01$ GeV, $m_\chi=0.001$ GeV and $m_N=0.00001$ GeV. Choosing such low masses for the heavy neutrinos fixes the Yukawa coupling to be $\mathcal{O}(10^{-11})$ in order to satisfy the sub-eV light neutrino masses. Furthermore, we have chosen the dark sector coupling to be $y_{DS}=10^{-8}$. In contrast to the high mass region, in this region the heavy neutrinos are not thermalized owing to the small Yukawa couplings. Hence, in this case the number densities of the two heavy neutrinos are also to be tracked in addition to the number densities of the dark sector particles, and the heavy neutrinos also contribute to the relic via freeze-in since their decay to SM particles is kinematically forbidden. The evolution of the number densities are dictated by the following Boltzmann equations:

\begin{equation}
\frac{dn_\chi^{FI}}{dt}+3Hn_\chi^{FI}=n_{\chi eq}n_{\phi eq}\langle\sigma v_{\chi\phi\rightarrow SMSM}\rangle+2n_\phi^{FI}\langle\Gamma_{\phi\rightarrow\chi N}\rangle,
\label{equ:reg3_1}
\end{equation}

\begin{equation}
\frac{dn_\phi^{FI}}{dt}+3Hn_\phi^{FI}=n_{\chi eq}n_{\phi eq}\langle\sigma v_{\chi\phi\rightarrow SMSM}\rangle-2n_\phi^{FI}\langle\Gamma_{\phi\rightarrow\chi N}\rangle,\,\textrm{and}
\label{equ:reg3_2}
\end{equation}

\begin{equation}
\frac{dn_{N_i}^{FI}}{dt}+3Hn_{N_i}^{FI}=n_{heq}\langle\Gamma_{h\rightarrow\nu N_i}\rangle+n_{Weq}\langle\Gamma_{W^\pm\rightarrow l^\pm N_i}\rangle+n_{Zeq}\langle\Gamma_{Z\rightarrow \nu N_i}\rangle+n_\phi^{FI}\langle\Gamma_{\phi\rightarrow\chi N}\rangle \  \  \ \ \ \forall \ i\in\{1,2\},
\end{equation}
 where the SM state pairs in equations \ref{equ:reg3_1} and \ref{equ:reg3_2} are $h\nu$, $Z\nu$, and $W^\pm l^\mp$. The DM particles are frozen-in after being produced from the annihilations and decays of the SM states. Between the two, it is the decay of the SM states that dominate the heavy neutrino production. Furthermore, decay of $\phi$ into $\chi$ and $N_i$ are also taken into account accordingly. The evolution equation for $\chi$ and $\phi$ produced from BH evaporation are the same as Eqns.~\ref{equ:reg1_3} and \ref{equ:reg1_4} while the one for $N_i$ is given by

\begin{equation}
\frac{dn_{N_i}^{BH}}{dt}+3Hn_{N_i}^{BH}=\frac{\rho_{BH}}{M}\frac{dN_{N_i}}{dt}+n_\phi^{BH}\langle\Gamma_{\phi\rightarrow\chi N}\rangle .
\end{equation}

\begin{figure}[h!]
\centering
\includegraphics[scale=0.55]{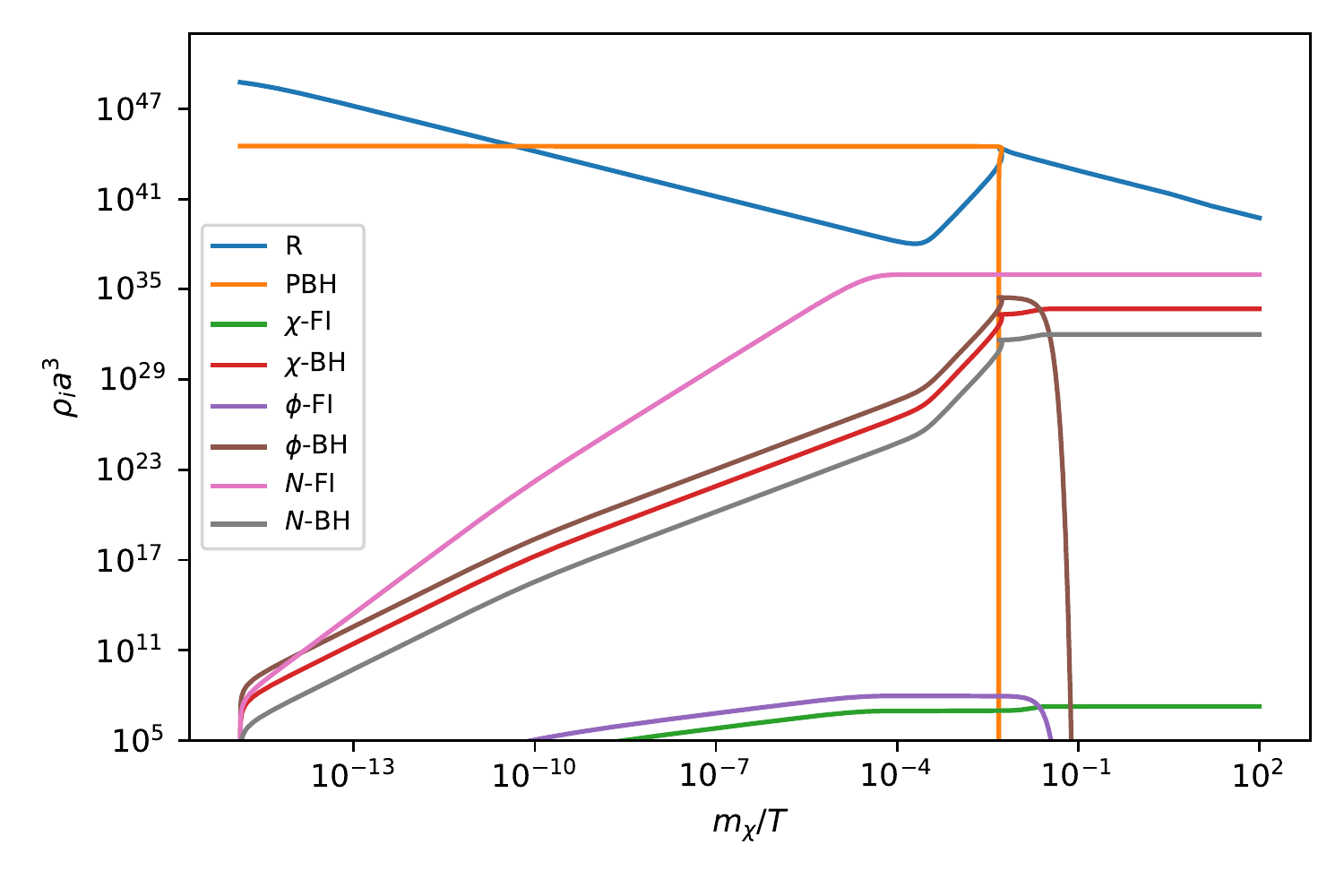}\includegraphics[scale=0.55]{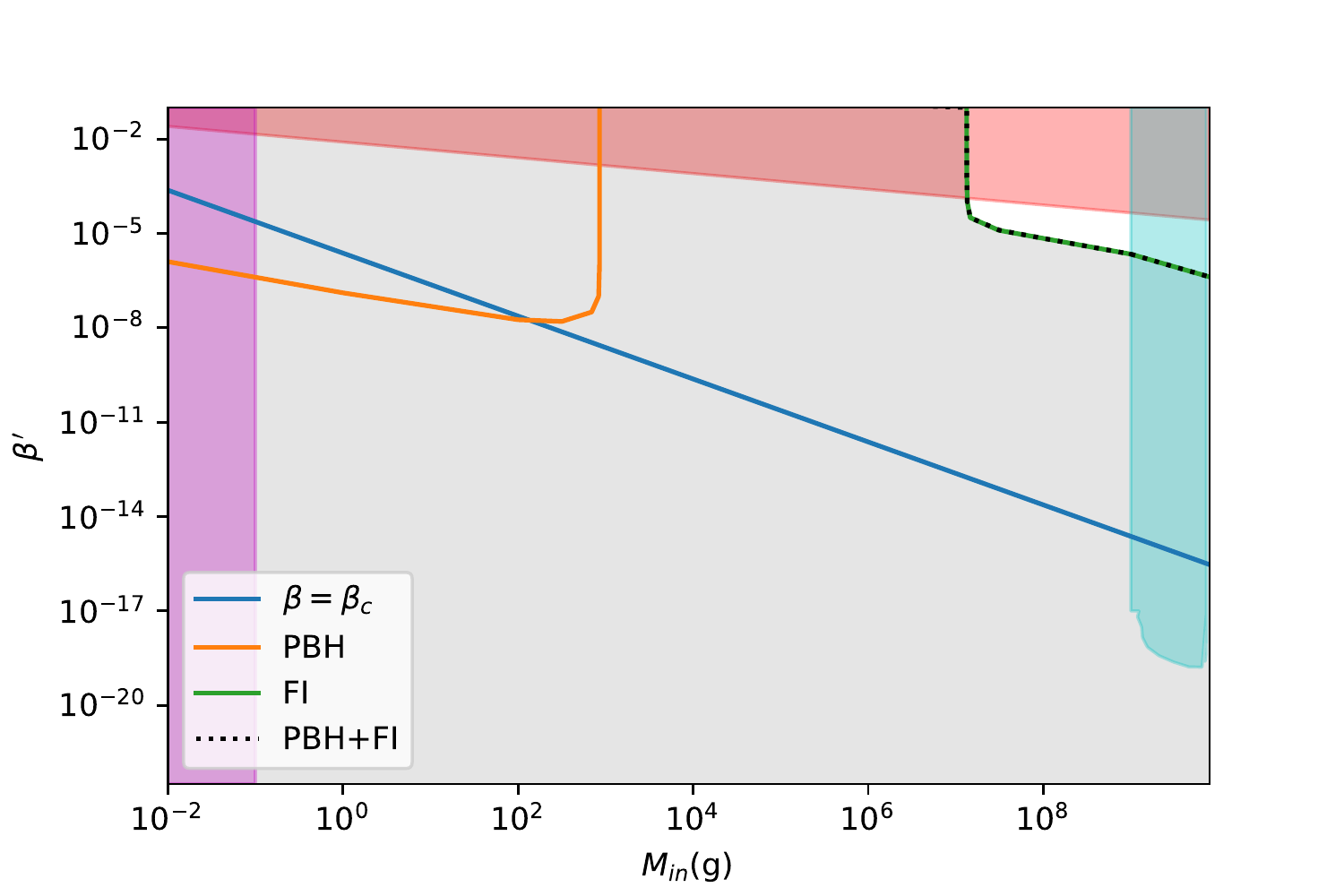}
\caption{ Evolution of various components of universe for benchmark points $M_{in}=10^{7.5}$ g and $\beta^\prime=10^{-4.9}$ where $\Omega h^2\sim0.12$ (left). Relic contours for Freeze-In (FI) and Primordial Black Hole (PBH) components individually as well as summed together along with various PBH constraints (right).}
\label{fig:relic3}
\end{figure}

Fig.~\ref{fig:relic3} (left) depicts the production of $\phi$, $\chi$, and $N_i$ and the subsequent decay of $\phi$ as the $\chi$ and $N_i$ form the dark matter relic. The co-moving energy densities of both generations of heavy neutrinos are calculated separately and their sum is displayed in the plot. The heavy neutrinos in this model can decay to light neutrinos through some highly off-shell mediators like the $Z$ boson. Such processes however take longer ($\sim 10^{19}$ sec) than the age of the universe ($\sim 10^{17}$ sec) and hence the heavy neutrinos are stable on cosmological time scale to be the major DM candidate. The sterile neutrinos in this context also respect the bounds from production of X-rays as calculated following \cite{Caputo:2019djj}.

\FloatBarrier

The relic contours are displayed in Fig.\ref{fig:relic3} (right) which shows that a small portion of parameter space is allowed (underabundant) in the PBH dominated region for this case. This unshaded region lies in the relatively higher $\beta^\prime$ and $M_{in}$ region of $\beta^\prime\sim(10^{-6}-10^{-2})$ and $M_{in}\sim(10^7-10^9)$ g. The constraints from non-cold DM cited in Sec.~\ref{subsec:case1} is much more relevant for low mass cases (case-III and case-IV). In both cases, we have chosen our benchmark points in such a way that it respects warm DM constraints.

\subsection{Case-IV: Low Mass Region with $\boldsymbol{m_N>m_\phi=m_\chi}$}

We now turn to the case where the $\phi$ and $\chi$ are degenerate ($m_\chi=m_\phi=0.001$ GeV) while still operating in the low mass regime. A $0.01$ GeV heavy neutrino pair restricts the relevant Yukawa couplings to $\mathcal{O}(10^{-10}-10^{-9})$ for a sub-eV light neutrino mass and thus the heavy neutrinos remain out of equilibrium. The heavy neutrinos are produced from out of equilibrium decays of $h$, $Z$, and $W^\pm$ and then decay further into $\phi$ and $\chi$ which form the dark matter relic. These statements are encapsulated in the following Boltzmann equations:

\begin{equation}
\frac{dn_\chi^{FI}}{dt}+3Hn_\chi^{FI}= (n_{N_1}^{FI}+n_{N_2}^{FI})\langle\Gamma_{N\rightarrow\chi \phi}\rangle,
\label{equ:reg4_1}
\end{equation}

\begin{equation}
\frac{dn_\phi^{FI}}{dt}+3Hn_\phi^{FI}=(n_{N_1}^{FI}+n_{N_2}^{FI})\langle\Gamma_{N\rightarrow\chi \phi}\rangle, \,\textrm{and}
\label{equ:reg4_2}
\end{equation}

\begin{equation}
\frac{dn_{N_i}^{FI}}{dt}+3Hn_{N_i}^{FI}=n_{heq}\langle\Gamma_{h\rightarrow\nu N_i}\rangle+n_{Weq}\langle\Gamma_{W^\pm\rightarrow l^\pm N_i}\rangle+n_{Zeq}\langle\Gamma_{Z\rightarrow \nu N_i}\rangle-n_{N_i}^{FI}\langle\Gamma_{N\rightarrow\chi\phi}\rangle \  \  \ \ \ \forall \ i\in\{1,2\}.
\label{equ:reg4_3}
\end{equation}

\begin{figure}[h!]
\centering
\includegraphics[scale=0.55]{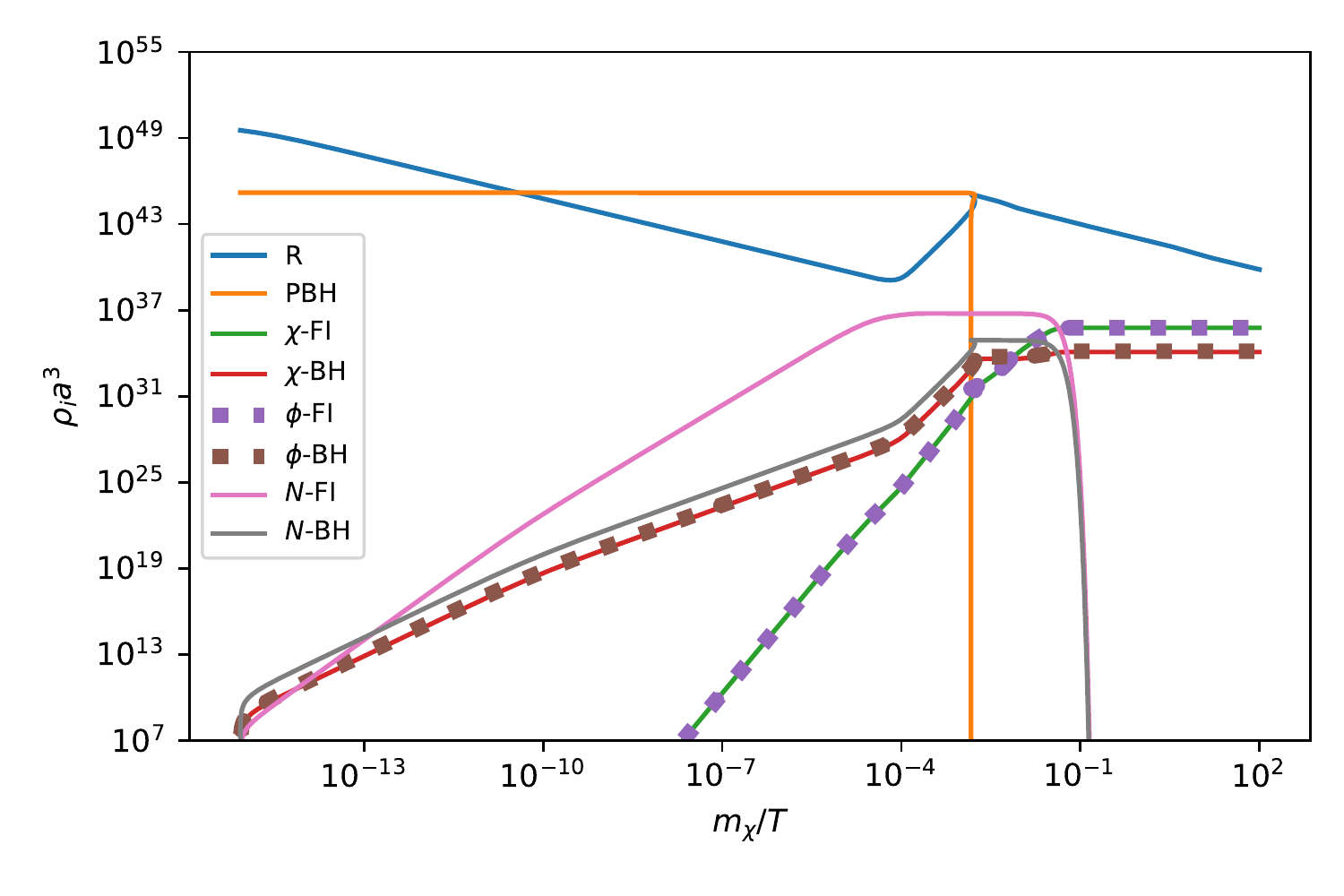}\includegraphics[scale=0.55]{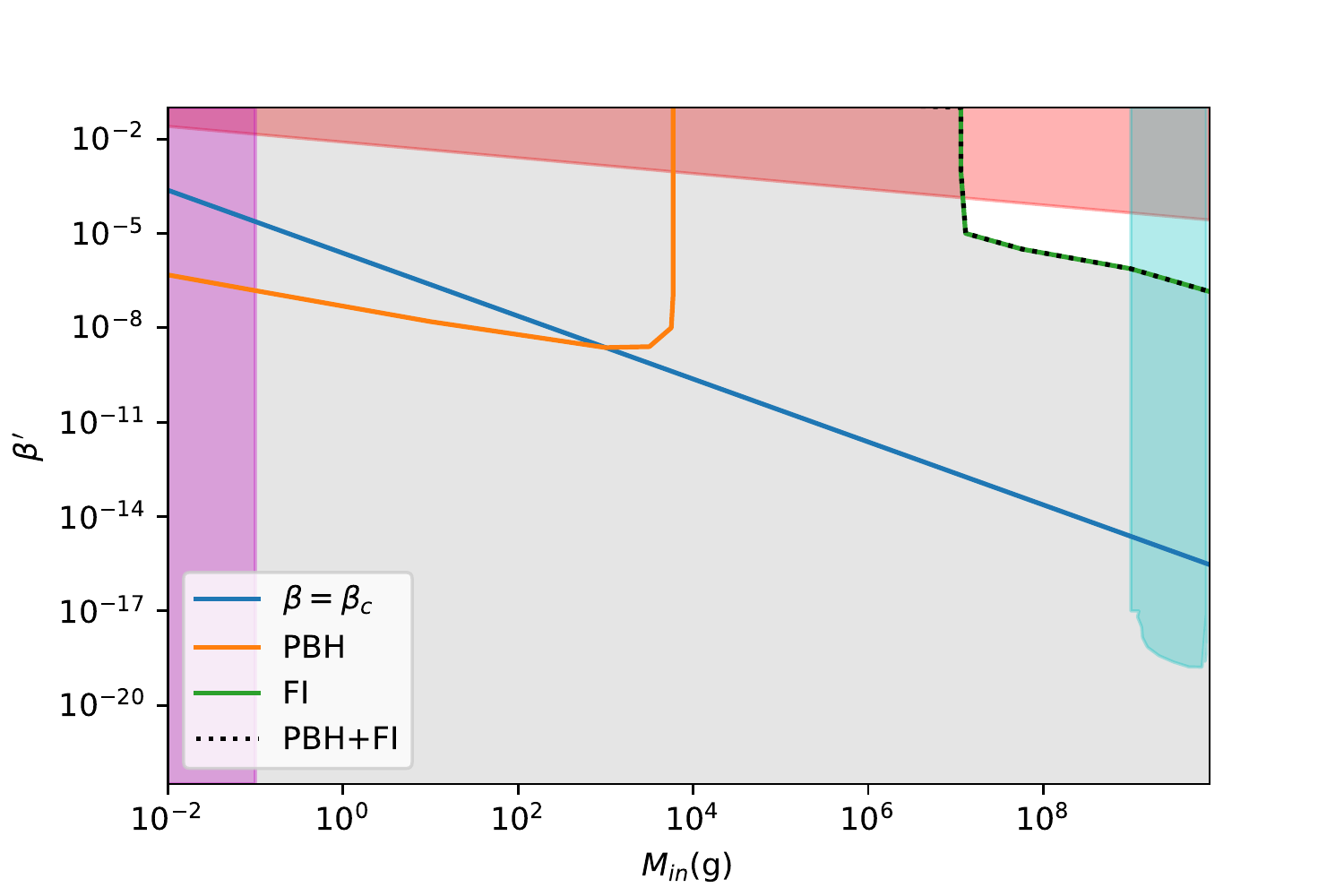}
\caption{ Evolution of various components of universe for benchmark points $M_{in}=10^{7.114}$ g and $\beta^\prime=10^{-5.0}$ where $\Omega h^2\sim0.12$ (left). Relic contours for Freeze-In (FI) and Primordial Black Hole (PBH) components individually as well as summed together along with various PBH constraints (right).}\label{fig:relic4}
\end{figure}

Also, the equations of their BH counterparts are given by the following:

\begin{equation}
\frac{dn_\chi^{BH}}{dt}+3Hn_\chi^{BH}=\frac{\rho_{BH}}{M}\frac{dN_{\chi}}{dt}+(n_{N_1}^{BH}+n_{N_2}^{BH})\langle\Gamma_{N\rightarrow\chi \phi}\rangle,
\label{equ:reg4_4}
\end{equation}

\begin{equation}
\frac{dn_\phi^{BH}}{dt}+3Hn_\phi^{BH}=\frac{\rho_{BH}}{M}\frac{dN_{\phi}}{dt} +(n_{N_1}^{BH}+n_{N_2}^{BH})\langle\Gamma_{N\rightarrow\chi \phi}\rangle,\,\textrm{and}
\label{equ:reg4_5}
\end{equation}

\begin{equation}
\frac{dn_{N_i}^{BH}}{dt}+3Hn_{N_i}^{BH}=\frac{\rho_{BH}}{M}\frac{dN_{N_i}}{dt} -n_{N_i}^{BH}\langle\Gamma_{N\rightarrow\chi\phi}\rangle \  \  \ \ \ \forall \ i\in\{1,2\}.
\label{equ:reg4_6}
\end{equation}
We note here that in addition to SM states, PBH evaporation produces $\phi$, $\chi$ and $N_i$ particles, only for the $N_i$ to decay into $\phi$ and $\chi$ which is encapsulated in the above equations.
Fig.~\ref{fig:relic4} (left) makes it evident that for the most part, the degenerate $\chi$ and $\phi$ follows each other's co-moving energy density. The heavy neutrinos produced from SM decays and BH evaporation are wiped out by their decay contributions to $\chi$ and $\phi$ which are the DM candidates in this scenario.

\FloatBarrier

The contour plot of Fig.~\ref{fig:relic4} (right) shows that freeze-in components dictate the relic of the PBH dominated area of the parameter space. Also, the relic contour for only the PBH evaporation has the same slope throughout the radiation dominated region since for the relevant parameter space of initial BH mass, the initial BH temperature is always higher than the DM mass.

To summarize this section, we note that in Case-I, $\chi$ is the DM candidate and in Case-III, the DM candidates are $\chi$ and $N_i$. For Cases II and IV, $\chi$ and $\phi$ together constitute the DM sector. In Cases I and II, the relic is dominated by the PBH evaporation contribution to DM in the PBH dominated region. However, in the radiation dominated region, there is scope for the freeze-in contribution to dictate the relic. In the PBH dominated region for the specific benchmark values of the DM masses in Cases I and II, the parameter space is under-abundant beyond $M_{in}=10^{7.8}$ g. However, more parameter space opens up in radiation dominated region. In the low mass scenarios (\textit{i.e.,} Cases III and IV), the relic is freeze-in dominated and the under-abundant region of parameter space is around $\beta^\prime\sim(10^{-6}-10^{-2})$ and $M_{in}\sim(10^7-10^9)$ g while the radiation dominated region has been rendered over-abundant in both these cases. In such freeze-in only production, the role of PBH is very important since it can help to dilute freeze-in relic through entropy injection for an otherwise overabundant scenario. Thus, the interplay of freeze-in and PBH production of dark matter offers scope for opening up different regions of parameter space which can account for the measured relic density.

%%%%%%%%%%%%%%%%%%%%%%%%%%%%%%%%%%%%%%%%%%%%%%%%%%%%%%%%%%%%%
\section{Conclusion}
\label{sec:conc}
Primordial black holes are interesting cosmological objects that can radiate all kinds of particles in a democratic fashion as they evaporate via Hawking radiation. The presence of a black hole in primordial times thus opens up a new avenue in the solution to the dark matter problem in addition to more traditional WIMP and FIMP production mechanisms. Very massive dark matter produced from the Hawking radiation of PBHs can be abundant enough to explain the observed relic on its own when the dark particle sector is completely isolated from the SM sector throughout the evolution of the universe. In this work however we have analysed the co-existence of both Hawking radiation as well as the freeze-in production of dark matter. For that purpose we have chosen a Littlest Seesaw model involving two heavy RHNs  along with an additional scalar $\phi$ and a fermion $\chi$ which are feebly coupled to the heavy RHNs. This extended BSM sector offers many interesting possibilities.

To understand the phenomenological importance of the model, we studied two very distinct mass regions each with two different mass hierarchies. The mass hierarchy plays a key role in determining the DM candidature along with the mass range that is considered. A very massive RHN thermalises with the cosmic bath due to high couplings but a sub-MeV RHN does not have a coupling strong enough to thermalise. Furthermore, an RHN of huge mass has decay channels via SM particles which is not exactly the case for a sub-MeV RHN. Apart from driving the velocity averaged scattering cross section for the relic, the role of the portal coupling in various decay processes as allowed kinematically by different choices of mass hierarchy is also a key feature in determining the DM phenomenology.

Solving a set of coupled Boltzmann equations simultaneously in each case to get an accurate description of the evolution of relevant species, we find that in the high mass region the PBH evaporation component of DM governs the relic completely in the PBH dominated region of parameter space. In the radiation dominated region however, the freeze-in component also determines the relic in certain regions of parameter space. In the low mass regions the end result is drastically different - the freeze-in component completely dictates the relic in the PBH dominated region while the radiation dominated region is completely disallowed by overabundance. Our results indicate that the study of multiple production mechanisms of DM remains an attractive field of study, and an interesting future possibility is to understand the DM phenomenology of the present model but with a rich spectrum of primordial black holes in the early universe.

\appendix
\section{Evolution Equations}
\label{sec:equations}
The evolution equations of radiation energy density ($\rho_R$) and temperature of universe ($T$) are given by the following equations \cite{Perez-Gonzalez:2020vnz},

\begin{equation}
\frac{d\rho_R}{dt}+4H\rho_R=-\frac{\varepsilon_{SM}(M)}{\varepsilon(M)}\frac{1}{M}\frac{dM}{dt}\rho_{BH}
\end{equation}

\begin{equation}
\frac{dT}{dt}=-\frac{T}{\Delta}\bigg(H+\frac{\varepsilon_{SM}(M)}{\varepsilon(M)}\frac{1}{M}\frac{dM}{dt}\frac{g_\star(T)}{g_{\star s}(T)}\frac{\rho_{BH}}{4\rho_R}\bigg)
\end{equation}
where, $\Delta=1+\frac{T}{3g_{\star s}(T)}\frac{dg_{\star s}(T)}{dT}$

\acknowledgments

BC acknowledges support from the Department of Science and Technology, India, under Grant CRG/2020/004171. SS wants to thank Yuber F. Perez-Gonzalez for useful discussions.

\bibliographystyle{apsrev}
\bibliography{bibliography}

\end{document}